\documentstyle[epsfig,12pt]{article}  
\newcommand{\bce}{\begin{center}}
\newcommand{\ece}{\end{center}}
\newcommand{\beq}{\begin{equation}}
\newcommand{\eeq}{\end{equation}}
\newcommand{\bea}{\vspace{0.25cm}\begin{eqnarray}}
\newcommand{\eea}{\end{eqnarray}}

\newcommand{\ba}{\begin{array}}
\newcommand{\ea}{\end{array}}

\newcommand{\r}{\mbox{{\boldmath
$\rho$}}}

\newcommand{\qb}{\mbox{{\bf
q}}}

\newcommand{\doublespace}{
\renewcommand{\baselinestretch}{1.6}\large\normalsize}

\def\lsim{\mathrel{\rlap{\lower4pt\hbox{\hskip1pt$\sim$}}
    \raise1pt\hbox{$<$}}}         
\def\gsim{\mathrel{\rlap{\lower4pt\hbox{\hskip1pt$\sim$}}
    \raise1pt\hbox{$>$}}}         

    \def\beq{\begin{equation}}
    \def\endeq{\end{equation}}
    \def\bea{\begin{eqnarray}}
    \def\arr{\begin{eqnarray}}
    \def\eea{\end{eqnarray}}

\def\q2{$Q^{2}$}
\def\s2{2$S$}

\begin{document}
\doublespace
\begin{titlepage}
\vspace*{-2cm}
\begin{flushright}
\bf
MPI-H-V11-1998\\
\vspace*{-.29cm}
March 1998\\
\end{flushright}
\bigskip

\begin{center}

  {\Large\bf
  Light-cone path integral approach to
  the
  Landau--Pomeranchuk--Migdal
  effect
  and the SLAC data on bremsstrahlung
  from
  high energy
  electrons
  }
  
  \vspace{0.3
  cm}
  
  {\large B.G. Zakharov
  }
  \bigskip

{\it
Max--Planck Institut f\"ur Kernphysik, Postfach
103980\\
69029 Heidelberg,
Germany\medskip\\
L. D. Landau Institute for Theoretical
Physics,
GSP-1, 117940,\\ ul. Kosygina 2, 117334 Moscow,
Russia
\medskip\\}

\vspace{.1cm}

{\bf\large
Abstract}
\end{center}
We analyze the recent data of the SLAC E-146
collaboration
\cite{SL2}
on
the Landau-Pomeranchuk-Migdal effect for bremsstrahlung
from
8 and 25 GeV
electrons
within a rigorous light-cone path
integral
approach previously developed in
Refs. \cite{LPM1,LPM2}.
Numerical calculations have been carried
out
treating rigorously the Coulomb effects
and
including the inelastic
processes.
Comparison with the experimental data is performed taking
into
account
multi-photon emission and photon absorption. For most of
the
targets our predictions are in excellent agreement with
the
experimental
data.

\vspace*{3cm}

\noindent

\end{titlepage}

\newpage
\section{Introduction}
Landau and Pomeranchuk \cite{LP} showed in 1953 that at high
energies
multiple scattering suppresses radiation process in
matter.
Within classical
electrodynamics
they obtained for a high energy
electron
the bremsstrahlung
spectrum
$\propto 1/\sqrt{k}$ ($k$ is the photon momentum) in contrast
to
the $1/k$ Bethe-Heitler spectrum for an isolated
atom.
Later this prediction was corroborated by Migdal \cite{Migdal}
who
developed a
quantum-mechanical
theory of bremsstrahlung and pair production in
medium.

Since the works by Landau and Pomeranchuk \cite{LP}
and
Migdal \cite{Migdal} suppression of the radiation processes
in
medium, called in the current literature
the
Landau-Pomeranchuk-Migdal (LPM) effect, has been attracting
much
attention (for a complete list of
references
see
\cite{TM,Ryazanov,Akh,Klein}).
The LPM
effect
has
been
qualitatively
corroborated in the Serpukhov experiment on bremsstrahlung
from
40 GeV electrons \cite{Serp} and in the
experiments
with cosmic rays \cite{C1,C2}. However, only recently the
first
quantitative measurement of the LPM effect
for
bremsstrahlung from 8 and 25 GeV electrons has been performed
by
the SLAC E-146 collaboration
\cite{SL1,SL2}.
This
experiment
stimulated a new theoretical
activity
\cite{ShF,BlanD,LPM1,B2,LPM2,Blan}
on the LPM effect in
QED.
Nevertheless, a
detailed
theoretical analysis of the SLAC data is still
lacking.

In the present paper we analyze the experimental
data
\cite{SL2} within a rigorous theory of the LPM effect
developed
in Ref. \cite{LPM1} (see also \cite{LPM2}) which is
applicable
for both QED and
QCD.
The approach of Ref. \cite{LPM1} is based
on
the light-cone path integral formalism in the
coordinate
representation and the technique of statistical
averaging
over the medium potential previously developed in
Ref. \cite{A2eBGZ}.
In Ref. \cite{LPM1} evaluation of the radiation rate has been
reduced
to solving a two-dimensional
Schr\"odinger
equation with an imaginary potential. In QED this
potential
is proportional to the dipole cross section for
scattering
of $e^{+}e^{-}$ pair off an
atom.
From the
conceptual
viewpoint the approach of Ref. \cite{LPM1} is equivalent
to
Migdal's analysis within time-ordered
perturbation
theory in the momentum representation. However, in Migdal's
formalism
a simple expression for the radiation rate can be
obtained
only within the Fokker-Planck approximation, in
which
the Coulomb effects are treated to logarithmic
accuracy.
It works well in the limit
of
strong LPM suppression in an infinite medium,
but
it is not good for real
situations
due to the uncertainty in the value of
the
Coulomb
logarithm.
Previously, in
Ref. \cite{LPM2},
we compared the theoretical predictions with
the
experimental spectrum for 25 GeV beam on a gold target with
thickness
$L=0.7\%\,X_{0}$ ($X_{0}$ is the radiation
length)
presented in the first SLAC publication
\cite{SL1}.
In the present paper
we
carry out a
detailed
comparison of the theoretical
predictions
with the
complete
data taken by the SLAC E-146 collaboration
\cite{SL2}
for 8 and 25 GeV
electrons
in a variety of materials. In our analysis we take into
account
multi-photon
emission which plays an important role for the targets
with
thickness $L\sim 2-6\%\,X_{0}$ used in the SLAC
experiment.
We also take into account photon
absorption.
As compared with the analysis of
Ref. \cite{LPM2}
we use a parametrization of the dipole cross section
which
is more accurate for heavy
elements.

The presentation is organized as follows. In section
2
we give our basic formulas for the radiation rate
obtained
neglecting multi-photon
emission.
In section 3 we discuss relationship
between
the probability distribution in the radiated energy, measured at
the
SLAC experiment, and the
probability
of one-photon emission. Including one- and two-photon emission
we
derive a simple relation between these
quantities.
In section 4 we present the numerical results for
LPM
suppression factors for a variety of materials and compare
the
theoretical predictions with the SLAC data. The results
are
summarized in section
5.

\section{Expression for the probability of photon emission}
In Ref. \cite{LPM1} we have expressed
the
cross
section for the radiation process $a\rightarrow bc$
through
the Green's
function
of a two-dimensional Schr\"odinger equation in impact-parameter
space
for which the longitudinal coordinate $z$ plays the role of
time.
This equation describes evolution of the light-cone wave
function
of a
fictitious
three-body $\bar{a}bc$ state. In this state the transverse
coordinate
of the particle $a$ coincides with the center-of-mass
of
the $bc$ system, and the only dynamic spatial variable
is
the transverse separation between the particles $b$ and
$c$. For
radiation of a photon with momentum $k$ from an
electron
with high energy $E_{e}\gg m_{e}$ ($m_{e}$ is the electron
mass)
incident
on an
amorphous
target
the corresponding Hamiltonian, describing
the
$e^{+}e^{-}\gamma$
system,
is given
by
\beq
{
H}=\frac{{\qb}^{2}}{2\mu(x)}+v(\r,z)\,,
\label{eq:10}
\eeq
\beq
v(\r,z)=-i\frac{n(z)\sigma(|\r
|x)}{2}\,,
\label{eq:20}
\eeq
where $x=k/E_{e}$ is the
photon
fractional longitudinal
momentum,
the Schr\"odinger mass
is
$\mu(x)=E_{e}x(1-x)$,
$n(z)$ is the number density of the target (assumed to
be
independent of the
transverse
coordinate),
and
$\sigma(\rho)$ is the dipole cross
section
for scattering of $e^{+}e^{-}$ pair of the transverse
size
$\rho$ off an
atom.
The
transverse
coordinate $\r$ in Eqs. (\ref{eq:10}),~(\ref{eq:20})
is
the transverse distance between electron and
photon
in the $e^{+}e^{-}\gamma$ system. In terms of $x$ and $\r$
the
electron-positron and photon-positron transverse separations
are
$\r_{e\bar{e}}=-x\r$ and $\r_{\gamma\bar{e}}=(1-x)\r$,
respectively.

Neglecting multi-photon
radiation,
the probability of photon
emission
can be written in the form
\cite{LPM1}
\beq
\frac{d P_{\gamma}}{d
x}=2\mbox{Re}
\int\limits_{-\infty}^{\infty} d
\xi_{1}
\int\limits_{\xi_{1}}^{\infty}d
\xi_{2}
\exp\left(-\frac{i\Delta\xi}{L_{f}}\right)
g(\xi_{1},\xi_{2},x)\left[{\cal
K}(0,\xi_{2}|0,\xi_{1})
-{\cal
K}_{v}(0,\xi_{2}|0,\xi_{1})
\right]\,.
\label{eq:30}
\eeq
Here ${\cal K}$ is the Green's function for the
Hamiltonian
(\ref{eq:10}), ${\cal K}_{v}$ is the vacuum Green's
function
for the Hamiltonian (\ref{eq:10}) with
$v(\r,z)=0$,
\beq
L_{f}=
\frac{2E_{e}(1-x)}{m_{e}^{ 2
}x}
\label{eq:40}
\eeq
is the so-called photon formation
length.
The vertex operator
$g(\xi_{1},\xi_{2},x)$
is given
by
\beq
g(\xi_{1},\xi_{2},x)=
\frac{\alpha[4-4x+2x^{2}]}{4x}\,
\qb(\xi_{2})\cdot\qb(\xi_{1})+
\frac{\alpha
m_{e}^{2}x}{2E_{e}^{2}(1-x)^{2}}\,\,,
\label{eq:50}
\eeq
where
$\alpha=1/137$.
The two terms on the right-hand side of
Eq. (\ref{eq:50})
correspond to the $e\rightarrow e'\gamma$
transitions
conserving and changing the electron
helicity.

Treating the potential (\ref{eq:20}) as a perturbation one
can
represent
the
radiation rate (\ref{eq:30}) in the following form
\cite{LPM2}
\beq
\frac{d P_{\gamma}}{d
x}=
\frac{d P_{\gamma}^{BH}}{d x}+\frac{d P_{\gamma}^{abs}}{d
x}\,,
\label{eq:60}
\eeq
\beq
\frac{d P_{\gamma}^{BH}}{d
x}=T
\frac{d \sigma^{BH}}{d
x}\,,
\label{eq:70}
\eeq
\beq
\frac{d \sigma^{BH}}{d
x}=
\int
d\r\,
W_{e}^{e\gamma}(x,\r)
\sigma(\rho
x)\,\,.
\label{eq:70p}
\eeq
$$
W_{e}^{e\gamma}(x,\r)=
\frac{1}{2}
\sum\limits_{\{\lambda_{i}\}}
|\Psi(x,\r,\{\lambda_{i}\})|^{2}\,,
$$
\bea
\frac{d P_{\gamma}^{abs}}{d
x}=-\frac{1}{4}\mbox{Re}
\sum\limits_{\{\lambda_{i}\}}
\int\limits_{0}^{L}
dz_{1}n(z_{1})
\int\limits_{z_{1}}^{L}
dz_{2}n(z_{2})
\int
d\r\,
\Psi^{*}(x,\r,\{\lambda_{i}\})\nonumber\\
\times\sigma(\rho
x)
\Phi(x,\r,\{\lambda_{i}\},z_{1},z_{2})
\exp\left[-\frac{i(z_{2}-z_{1})}{L_{f}}\right]\,.
\label{eq:80}
\eea
Here
$
T=\int_{0}^{L} dz
n(z)
$
is the optical thickness of the target (we assume that $n(z)=0$ at
$z<0$
and
$z>L$),
$\Psi(x,\r,\{\lambda_{i}\})$
is the light-cone wave function for the
transition
$e\rightarrow e'\gamma$, $\{\lambda_{i}\}$ is the set
of
the helicity
variables.
In Eq. (\ref{eq:80}) the
function
$\Phi(x,\r,\{\lambda_{i}\},z_{1},z_{2})$ is
the
solution
of
the
two-dimensional
Schr\"odinger
equation
\beq
i\frac{\partial\Phi(x,\r,\{\lambda_{i}\},z_{1},z_{2})}{\partial
z_{2}}=
{H}\Phi(x,\r,\{\lambda_{i}\},z_{1},z_{2})
\label{eq:90}
\eeq
with the Hamiltonian
(\ref{eq:10}).
The
boundary
condition for
$\Phi(x,\r,\{\lambda_{i}\},z_{1},z_{1})$
is
\beq
\Phi(x,\r,\{\lambda_{i}\},z_{1},z_{1})=
\Psi(x,\r,\{\lambda_{i}\})\sigma(\rho
x)\,.
\label{eq:100}
\eeq

The
light-cone
wave
function
$
\Psi(x,\r,\lambda_{e},\lambda_{e'},\lambda_{\gamma})
$
for
$\lambda_{e'}=\lambda_{e}$
is given
by
\bea
\Psi(x,\r,\lambda_{e},\lambda_{e'},\lambda_{\gamma})=
\frac{1}{2\pi}\sqrt{\frac{\alpha
x}{2}}
\left[\lambda_{\gamma}(2-x)+2\lambda_{e}x\right]
\exp(-i\lambda_{\gamma}\varphi)m_{e}K_{1}(\rho
m_{e}x)\,,
\label{eq:110}
\eea
for $\lambda_{e'}=-\lambda_{e}$ the only
nonzero
component is the one with
$\lambda_{\gamma}=2\lambda_{e}$
\bea
\Psi(x,\r,\lambda_{e},-\lambda_{e},2\lambda_{e})=
\frac{-i}{2\pi}\sqrt{2\alpha x^{3}}m_{e}K_{0}(\rho
m_{e}x)\,,
\label{eq:120}
\eea
where $\varphi$ is the azimuthal angle, $K_{0}$
and
$K_{1}$ are the modified Bessel
functions.

The first term on the right-hand side of
Eq.~(\ref{eq:60}),
coming from the term $\propto v$ in expansion of
the
Green's function ${\cal K}$ in the
potential,
corresponds to the
impulse
approximation. From this follows that
Eq.~(\ref{eq:70p})
gives
the
Bethe-Heitler cross
section.
Note that this
representation
for the Bethe-Heitler cross section can be also
derived
directly for bremsstrahlung on an isolated atom
using
the light-cone approach developed in Ref. \cite{NPZ},
where
the heavy quark production was
discussed.
In
Eq.~(\ref{eq:60})
LPM
suppression
is described by the second term which is analogous to
the
Glauber absorptive correction in hadron-nucleus
collisions.
In the
present
paper we will use for numerical calculations
representation
of the radiation rate in
form
(\ref{eq:60}). It allows one to
bypass
evaluation of the singular Green's function. This makes
it
convenient for accurate
computations
with a rigorous treatment of the Coulomb
effects.

In the limit $L_{f}\rightarrow 0$ the second term in
Eq.
(\ref{eq:60}) vanishes and the Bethe-Heitler regime
obtains.
A simple expression for the radiation rate can be
also
obtained in the limit $L_{f}\gg
L$.
One can easily show that in this regime the kinetic term in
the
Hamiltonian (\ref{eq:10}) can be neglected. This means
that
the transverse variable $\r$ is
approximately
frozen, and
$\Phi(x,\r,\{\lambda_{i}\},z_{1},z_{2})$
can be
written
in the eikonal
form
\beq
\Phi(x,\r,\{\lambda_{i}\},z_{1},z_{2})\approx
\exp
\left[-\frac{\sigma(\rho x)}{2}\int\limits_{z_{1}}^{z_{2}}dz
n(z)
\right]
\Psi(x,\r,\{\lambda_{i}\})\sigma(\rho
x)\,.
\label{eq:130}
\eeq
Using Eq. (\ref{eq:130}) one can easily obtain from
Eqs.
(\ref{eq:60}),~(\ref{eq:70}),~(\ref{eq:70p}) and (\ref{eq:80}) for
the
radiation rate in the frozen-size
approximation
\beq
\frac{dP_{\gamma}^{fr}}{dx}=
2\int d\r
\,W_{e}^{e\gamma}(x,\r)
\left\{1-\exp\left[-\frac{\sigma(\rho
x)T}{2}\right]\right\}\,.
\label{eq:140}
\eeq
Note that Eq. (\ref{eq:140}) is analogous to the formula for
the
cross
section for heavy quark production in hadron-nucleus
collision
obtained in
Ref. \cite{NPZ}.

In the momentum space the spectrum
(\ref{eq:140})
after the Fourier transform can be rewritten in the
form
\beq
\frac{dP_{\gamma}^{fr}}{dx}=
\int
d\qb\,P(x,\qb)I(\qb)\,,
\label{eq:150}
\eeq
where
$$
I(\qb)=\frac{1}{(2\pi)^{2}}\int
d\r
\exp\left[i\qb\r
-\frac{\sigma(\rho)T}{2}\right]
$$
is the distribution function in momentum transfer for
the
electron
after passing through the target \cite{A2eBGZ}, and the
function
$$
P(x,\qb)=
\int d\r
\,W_{e}^{e\gamma}(x,\r)
\left[1-\exp(i\qb\r
x)\right]
$$
describes the probability of photon emission for
scattering
of the electron with momentum transfer equals
$\qb$.
The factorized form of the integrand in
Eq.~(\ref{eq:150})
reflects
the fact that for the photons with $L_{f}\gg L$ the target acts as
a
single
radiator.
In the limit $x\rightarrow 0$ from Eq.~(\ref{eq:150}) one
can
obtain the spectrum of Ref. \cite{ShF} evaluated within
classical
electrodynamics.

The dipole cross section,
entering
the imaginary potential (\ref{eq:20}),
for
an atom with the atomic number $Z$ can
be
written in the
form
\beq
\sigma(\rho)=\rho^{2}C(\rho)\,,
\label{eq:160m}
\eeq
\beq
C(\rho)=Z^{2}C_{el}(\rho)+Z
C_{in}(\rho)\,.
\label{eq:160}
\eeq
In Eq.~(\ref{eq:160}) the terms $\propto Z^{2}$ and $\propto
Z$
correspond to elastic and inelastic intermediate states in
interaction
of $e^{+}e^{-}$ pair with an
atom.
The components of the light-cone wave
function
$\Psi(x,\r,\{\lambda_{i}\})$ defined
by
Eqs.~(\ref{eq:110}),~(\ref{eq:120}) decrease steeply
at
$|\r|\gsim 1/m_{e}x$. As a consequence, the
Bethe-Heitler
cross
section
(\ref{eq:70p}) is dominated by the region $\rho \lsim
1/m_{e}x$.
For (\ref{eq:80}) the dominating values of $\rho$ are even
smaller
due to absorption of the large-size
configurations.
For this reason the bremsstrahlung
rate
is only
sensitive
to the behavior of $\sigma(\rho)$ at $\rho\lsim 1/m_{e}\ll
a$,
here $a\sim r_{B}Z^{-1/3}$ is the atomic
size.
In this region
both
the $C_{el}$ and $C_{in}$ have only weak logarithmic
dependence
on $\rho$, and we
can
parametrize them in the
form
\bea
C_{el}(\rho)=
4\pi\alpha^{2}
\left[\log\left(\frac{2a_{el}}{\rho}\right)+\frac{(1-2\gamma)}{2}
-f(Z\alpha)\right]\,,
\label{eq:170}
\eea
\beq
f(y)=y^{2}\sum\limits_{n=1}^{\infty}
\frac{1}{n(n^{2}+y^{2})}\,\,,
\label{eq:171}
\eeq
\bea
C_{in}(\rho)=
4\pi\alpha^{2}
\left[\log\left(\frac{2a_{in}}{\rho}\right)+\frac{(1-2\gamma)}{2}
\right]\,,
\label{eq:180}
\eea
where $\gamma=0.577$ is Euler's
constant.
Eq.~(\ref{eq:170}) defines $C_{el}(\rho)$ for $\rho\gsim
R_{A}$,
here $R_{A}$ is the nucleus radius. At $\rho\lsim
R_{A}$
on the right-hand side of Eq.~(\ref{eq:170}) $\rho$ must
be
replaced by
$R_{A}$.
The parametrization (\ref{eq:170}) of the elastic
component
corresponds to the result of calculation of $C(\rho)$ for
scattering
of
$e^{+}e^{-}$
pair on the atomic potential $\phi(r)=(Ze/4\pi
r)\exp(-r/a_{el})$.
The first two terms in the square brackets on
the
right-hand side of Eq. (\ref{eq:170}) give the
contribution
of the Born approximation while the last
one
is related to the Coulomb correction. It is expressed
through
function (\ref{eq:171}) which was introduced in
the
well-known analysis of pair production and
bremsstrahlung
by Davies, Bethe and Maximon
\cite{Coulomb}.
This correction becomes important only for heavy
elements. For
$Z\sim 70-90$ it decreases $C_{el}(\rho\sim 1/m_{e})$,
and
accordingly the Bethe-Heitler cross section, by $\sim
7-10\%$.
Using the parametrizations (\ref{eq:170}),
(\ref{eq:180})
and formulas (\ref{eq:110}),~(\ref{eq:120}) for
the
light-cone wave function one can obtain
from
Eq.~(\ref{eq:70p}) for the
cross
section of photon emission on an isolated
atom
\beq
\frac{d\sigma^{BH}}{dx}=
\frac{d\sigma^{BH}_{nf}}{dx}+
\frac{d\sigma^{BH}_{sf}}{dx}\,,
\label{eq:190}
\eeq
\beq
\frac{d\sigma^{BH}_{nf}}{dx}=
\frac{4\alpha^{3}(4-4x+x^{2})}{3
m_{e}^{2}x}
\left\{Z^{2}[F_{el}-f(Z\alpha)]+
Z
F_{in}+\frac{Z(Z+1)}{12}
\right\}\,,
\label{eq:200}
\eeq
\beq
\frac{d\sigma^{BH}_{sf}}{dx}=
\frac{4\alpha^{3}x}{3
m_{e}^{2}}
\left\{Z^{2}[F_{el}-f(Z\alpha)]+
Z
F_{in}-\frac{Z(Z+1)}{6}
\right\}\,,
\label{eq:210}
\eeq
where
$
F_{i}=\log(a_{i}m_{e}\exp(1/2))\,.
$
The two terms in Eq.~(\ref{eq:190}) correspond to
transitions
conserving (nf) and changing (sf) the electron
helicity.
We have adjusted $a_{el}$ and
$a_{in}$
to reproduce
the
Bethe-Heitler cross section
evaluated
in the standard approach with realistic atomic
formfactors
$F_{el}\approx
\log(184/Z^{1/3})\,,$
and $F_{in}\approx
\log(1194/Z^{2/3})\,,$
obtained within the Thomas-Fermi-Molier model
\cite{Tsai}.
This
gives
$a_{el}=0.81$ $ r_{B}Z^{-1/3}$
and
$a_{in}=5.3$
$r_{B}Z^{-2/3}$.

The strength of the LPM effect can be characterized by
the
suppression factor, $S(k,L)$, defined as (hereafter we assume
that
the target is
homogeneous)
\beq
S(k,L)=\frac{dP_{\gamma}}{dx}
\left(nL\frac{d\sigma^{BH}}{dx}\right)^{-1}\,.
\label{eq:220}
\eeq
In terms of the suppression factors $S_{nf}$ and $S_{sf}$,
defined
by relations analogous to Eq.~(\ref{eq:220}) for
transitions
conserving and changing the electron
helicity,
$S(k,L)$ is given
by
\beq
S(k,L)=
\left[
\frac{d\sigma^{BH}_{nf}}{dx}S_{nf}(k,L)+
\frac{d\sigma^{BH}_{sf}}{dx}S_{sf}(k,L)
\right]
\left(\frac{d\sigma^{BH}}{dx}\right)^{-1}\,.
\label{eq:230}
\eeq
In the kinematical domain of the SLAC experiment
\cite{SL2}
$x\ll 1$, and the spin-flip transitions give a negligible
contribution
to the radiation rate. As a result, $S(k,L)$ turns out to be
very
close to
$S_{nf}(k,L)\,$.

The edge effects vanish, and $S(k,L)$ becomes close to
the
suppression factor for infinite medium,
$S^{inf}(k)$,
for sufficiently large target thickness (or small
$x$)
$L\gg
L_{f}^{'}$,
here $L_{f}^{'}$ is the medium-modified photon formation
length.
In terms of the
representation
(\ref{eq:30}) $L_{f}^{'}$
is
the typical value of $|\xi_{2}-\xi_{1}|$ dominating the
integral
on the right-hand side of
Eq.~(\ref{eq:30}).
The medium-modified formation
length
characterizes
the
nonlocality of photon emission connected with
interference
effects. It is
important
from the viewpoint of applicability limits for the
probabilistic
treatment
of
multi-photon effects which will be discussed in the next
section.

To get an idea about the value of $L_{f}^{'}$
and
the strength of LPM suppression one can use the
results
of evaluation of the radiation rate in the oscillator
approximation
\cite{LPM1}. It corresponds to replacement of the
function
$C(\rho)$ (\ref{eq:160}) by its value
at
$\rho\sim\rho_{eff}x$, where $\rho_{eff}$ is the
typical
photon-electron separation dominating
the
Green's function ${\cal K}$ (the path integral
representation
is assumed) in
Eq.~(\ref{eq:30}).
For the case of not very strong LPM suppression, which will
be
interesting for analysis of the SLAC data, one can
take
$\rho_{eff}\sim 1/m_{e}x$. Then, the
Hamiltonian
(\ref{eq:10}) takes the oscillator form with the
frequency
$$
\Omega
=\frac{(1-i)}{\sqrt{2}}
\left(\frac{n
C_{osc}x}{E_{e}(1-x)}\right)^{1/2}
\,\,,
$$
where
$$
C_{osc}=C(\rho\sim
1/m_{e})\approx
4\pi\alpha^{2}[Z^{2}(F_{el}-f(Z\alpha)]+Z
F_{in}]\,.
$$
In this approximation the radiation rate in form
(\ref{eq:30})
can be
evaluated
using the known oscillator Green's
function.
For an infinite medium the oscillator
model
suppression factors depend on the dimensionless parameter
\cite{LPM1}
\bea
\eta=L_{f}|\Omega|=
2\left[\frac{nE_{e}(1-x)C_{osc}}{m_{e}^{4}x}\right]^{1/2}
\,\,\,\,\,\,\,\,\,\,\,\,\,\,\,\,\,\,\,\,\,\,\,\,\,\,\,
\nonumber
\\
\approx\left\{\frac{1.3
E_{e}^{2}(\mbox{GeV})
[1-10^{-3}k(\mbox{MeV})/E_{e}(\mbox{GeV})]}
{k(\mbox{MeV})X_{0}(\mbox{mm})}\right\}^{1/2}\,.
\label{eq:232}
\eea
Here we expressed $C_{osc}$ through
the
radiation length defined as in
Ref. \cite{Tsai}.
For weak
suppression
($\eta\ll 1$) $S_{nf}^{inf}\approx
1-16\eta^{4}/21$,
$S_{sf}^{inf}\approx 1-31\eta^{4}/21$, and for regime of
strong
suppression ($\eta \gg 1$) $S_{nf}^{inf}\approx
3/\eta\sqrt{2}$,
$S_{sf}^{inf}\approx 3\pi/2\eta^{2}$
\cite{LPM1}.
In terms of the photon momentum LPM suppression
becomes
significant
for
$k\lsim k_{LPM}$ where $k_{LPM}$, corresponding to
$\eta=1$,
is given
by
\beq
k_{LPM}\approx
\frac{1.3
E_{e}^{2}(\mbox{GeV})}{X_{0}(\mbox{mm})
[1+0.0013E_{e}(\mbox{GeV})/X_{0}(\mbox{mm})]
}\,.
\label{eq:233p}
\eeq
In Table 1 we give the values of $k_{LPM}$ for
the
target materials and electron energies used in the SLAC
experiment
\cite{SL2}. We are also give
in
this table the radiation
lengths.

Closer inspection of the expression
(\ref{eq:30})
in the
oscillator
approximation allows one to obtain the
following
estimate
for
the medium-modified photon formation length for the
transitions
conserving the electron
helicity
$
L_{f}^{'}\sim
L_{f}/\mbox{max}(1,\eta)
$
\cite{LPM1}. Using Eqs.~(\ref{eq:40}),~(\ref{eq:232})
one
can rewrite it
at
$x \ll
1$,
which will subsequently be interesting for analysis
of
the SLAC data \cite{SL2}, in the
form
\beq
L_{f}^{'}(\mbox{mm})\sim
10^{-3}\cdot
\mbox{min}\left\{
\frac{1.5
E_{e}^{2}(\mbox{GeV})}{k(\mbox{MeV})}\,,
1.32 E_{e}(\mbox{GeV})
\sqrt{\frac{X_{0}(\mbox{mm})}{k(\mbox{MeV})}}
\right\}\,.
\label{eq:234}
\eeq
We will use this approximate formula for estimate
of
$L_{f}^{'}$
for the SLAC
data.

\section{Multi-photon emission and
the probability distribution in the radiated
energy}

The experimental spectra of Ref. \cite{SL2} were
obtained
by measuring in a calorimeter the total
energy
of the photons radiated by the
electron.
This means that, up to a small correction connected with
photon
absorption, the spectra of
Ref. \cite{SL2}
correspond to the probability distribution in the electron
energy
loss,
${dP_{e}}/{dx}$, called in the literature the electron
struggling
function.
For sufficiently thin targets,
when
multi-photon effects can be neglected, the electron
struggling
function
is close
to
the probability of emission of a single
photon,
considered in previous
section.
For the SLAC data \cite{SL2} multi-photon emission is
small
for the gold target with $L=0.7\%\,X_{0}$,
but
for other targets with $L\sim 2-6\%\,X_{0}$
it
becomes
important.
A rigorous quantum-mechanical analysis of the LPM effect including
the
multi-photon
radiation in the kinematical region where $L_{f}^{'}\gsim
L$
is a complicated task requiring evaluation of higher order
diagrams.
We will compare the theoretical prediction with
the
data of Ref. \cite{SL2} for $k>5$ MeV (here $k$ is viewed as the
total
radiated
energy).
In this region of $k$ for the targets with $L\sim
2-6\%\,X_{0}$
the medium-modified photon
formation
length for 8 and 25 GeV
electrons
used in the SLAC experiment
turns
out to be considerably smaller than the
target
thickness.
At $L_{f}^{'} \ll L$ one can neglect the edge effects
and
the nonlocality of photon
emission.
This allows
us
to use the probabilistic approach to the multi-photon effects
in
which the probability of photon emission from an electron
per
unit length can be written in terms of the
medium-modified
Bethe-Heitler cross
section
\beq
\frac{d
\sigma^{eff}}{dx}=
S^{inf}(k=xE)\frac{d
\sigma^{BH}}{dx}\,.
\label{eq:240}
\eeq
Then, the electron struggling
function
can be obtained by
solving
the standard diffusion equation (see for instance
\cite{Rossi,Eyges}).
However, for analysis of the SLAC data we
need
${dP_{e}}/{dx}$ only at small $x$ and
for
sufficiently small target thicknesses $L\lsim
6\%\,X_{0}$.
In this case we can neglect the effect of the
electron
energy
loss on the probability of photon emission and restrict ourselves
to
the one- and two-photon processes. This allows us to
bypass
solving the diffusion equation and to
write
the electron struggling function in
the
following simple
form
\bea
\frac{dP_{e}}{dx}=
\int\limits_{0}^{L}
dz
U(0,z)n\frac{d
\sigma^{eff}}{dx}U(z,L)\hspace{6.5cm}
\nonumber\\
+
\int\limits_{0}^{L} dz_{1}\int\limits_{z_{1}}^{L}
dz_{2}
\int\limits_{x_{min}}^{x} dx_{1}dx_{2}
\delta(x_{1}+x_{2}-x)
U(0,z_{1})n\frac{d
\sigma^{eff}}{dx_{1}}U(z_{1},z_{2})
n\frac{d
\sigma^{eff}}{dx_{2}}U(z_{2},L)\,,
\label{eq:250}
\eea
where
\beq
U(z_{1},z_{2})=
\exp\left
[-(z_{2}-z_{1})\,n\!\int\limits_{x_{min}}^{1}dx_{1}
\frac{d
\sigma^{eff}}{dx_{1}}
\right]
\label{eq:260}
\eeq
is the attenuation factor for propagation of the
electron
in the target from $z_{1}$ to $z_{2}$ without photon
emission.
In Eqs.~(\ref{eq:250}), (\ref{eq:260}) we introduced an
infrared
regulator
$x_{min}$
which can be chosen from the condition $L_{f}^{'}(x_{min})\sim
L$.
This suggests that the electron struggling
function
(\ref{eq:250}) includes the
processes
with an arbitrary number of soft photons with $x\lsim
x_{min}$
radiated from initial and final electrons. As will be seen
later,
our final expression for the electron struggling
function,
similarly to the solution of the diffusion equation
\cite{Rossi},
is infrared stable, and $x_{min}$ will be set equal to
zero.
Note that within the light-cone path integral approach to
the
LPM effect of Ref. \cite{LPM1} the attenuation factor
(\ref{eq:260})
emerges as a result of evaluation of the radiative
corrections
to
the
transverse electron propagator connected with the
chain
diagrams
within the dilute gas approximation. This
approximation
corresponds neglecting the space overlapping of
different
electron-photon loops, {\it i.e.}, it gives the
attenuation
factor to leading (zeroth) order in
$L_{f}^{'}/L$.

Neglecting small surface effects we can rewrite
Eq. (\ref{eq:250})
in terms of the probability of photon emission evaluated
neglecting
multi-photon effects
as
\beq
\frac{dP_{e}}{dx}=
\exp\left
[-\int\limits_{x_{min}}^{1}dx_{1}
\frac{dP_{\gamma}}{dx_{1}}\right]
\left\{
\frac{dP_{\gamma}}{dx}
+
\frac{1}{2}
\int\limits_{x_{min}}^{x} dx_{1}dx_{2}
\delta(x_{1}+x_{2}-x)
\frac{dP_{\gamma}}{dx_{1}}
\frac{dP_{\gamma}}{dx_{2}}
\right\}\,.
\label{eq:270}
\eeq
For $L\ll X_{0}$ the exponential factor on
the
right-hand side of Eq.~(\ref{eq:270}) can be written
as
\beq
\exp\left
[-\int\limits_{x_{min}}^{1}dx_{1}
\frac{dP_{\gamma}}{dx_{1}}\right]
\approx
\exp\left
[-\int\limits_{x}^{1}dx_{1}
\frac{dP_{\gamma}}{dx_{1}}\right]
\left\{1-
\int\limits_{x_{min}}^{x}dx_{1}
\frac{dP_{\gamma}}{dx_{1}}
\right\}\,.
\label{eq:280}
\eeq
Then, using Eqs.~(\ref{eq:270}) and (\ref{eq:280}) we
obtain
\beq
\frac{dP_{e}}{dx}=
\frac{dP_{\gamma}}{dx}K(x)\,,
\label{eq:290}
\eeq
where
\bea
K(x)=
\exp\left
[-\int\limits_{x}^{1}dx_{1}
\frac{dP_{\gamma}}{dx_{1}}\right]\hspace{6.5cm}
\nonumber\\
\times
\left
\{1-
\int\limits_{x_{min}}^{x}dx_{1}
\frac{dP_{\gamma}}{dx_{1}}
+\frac{1}{2}
\int\limits_{x_{min}}^{x} dx_{1}dx_{2}
\delta(x_{1}+x_{2}-x)
\frac{dP_{\gamma}}{dx_{1}}
\frac{dP_{\gamma}}{dx_{2}}
\left(
\frac{dP_{\gamma}(x)}{dx}
\right)^{-1}
\right\}\,.
\label{eq:300}
\eea
We see that, as was
said
above, the electron struggling function defined
by
Eqs.~(\ref{eq:290}),~(\ref{eq:300})
is an
infrared
stable quantity. Therefore, we can set
$x_{min}=0$
and rewrite the multi-photon $K$-factor
(\ref{eq:300})
in the
form
\bea
K(x)=
\exp\left
[-\int\limits_{x}^{1}dx_{1}
\frac{dP_{\gamma}}{dx_{1}}\right]\hspace{5cm}
\nonumber\\
\times
\left
\{1-\frac{1}{2}
\int\limits_{0}^{x}dx_{1}\left[
\frac{dP_{\gamma}}{dx_{1}}+\frac{dP_{\gamma}}{dx_{2}}-
\frac{dP_{\gamma}}{dx_{1}}
\frac{dP_{\gamma}}{dx_{2}}
\left(
\frac{dP_{\gamma}(x)}{dx}
\right)^{-1}\right]
\right\}\,,
\label{eq:310}
\eea
where
$x_{2}=x-x_{1}$.
The major $x$-dependence of the $K$-factor (\ref{eq:310})
comes
from the exponential factor which reflects a simple fact
that
emission of the photons with the fractional momentum bigger than
$x$
is forbidden. We checked the accuracy of the relation
(\ref{eq:310})
using as a test solution the exact expression for
the
electron struggling
function
\beq
\frac{dP_{e}}{dx}=
\left[\log\left(\frac{1}{1-x}\right)\right]^{bt-1}
\Gamma(bt)^{-1}
\label{eq:320}
\eeq
(here $t=L/X_{0}$, and $\Gamma$ is the Euler
Gamma-function)
obtained
by Bethe and Heitler \cite{BHstrug} (see also
\cite{Eyges})
for the
model
bremsstrahlung cross
section
\beq
\frac{d\sigma}{dx}=b\left[\log\left(
\frac{1}{1-x}\right)\right]^{-1}\,.
\label{eq:330}
\eeq
This theoretical
experiment
shows that for the kinematical domain $5<k<500$ MeV,
which
will subsequently be
interesting,
Eq. (\ref{eq:290}) with the $K$-factor (\ref{eq:310})
has
inaccuracy $\lsim
0.5\%$.

The $K$-factor (\ref{eq:310}) was obtained ignoring the
edge
effects.
In principle, in the probabilistic approach one
can
obtain a formula for the
$K$-factor
including the
boundary
radiation. However, as was above mentioned, the probabilistic
approach
itself is
justified
only to zeroth order in $L_{f}^{'}/L$. On the other hand,
the
boundary radiation is an effect of the order of $\sim
L_{f}^{'}/L$.
For this reason an evaluation of
the
$K$-factor including the edge effects does not make much sense,
and
we will compare our predictions with experiment
using
Eqs.~(\ref{eq:290}),~(\ref{eq:310}).
For the targets with $L\sim 2-6\%\,X_{0}$ at $E_{e}=25$ GeV
the
corresponding errors
cannot
exceed a few percent at $k\sim 5-10$ MeV, and become negligible
for
$k\gsim 20-30$ MeV. For $E_{e}=8$ GeV they are negligible in
the
whole range of
$k$.

Due to the possibility of absorption of
the
radiated photons in the
target
the spectra of Ref. \cite{SL2} do not correspond exactly
to
the electron struggling
function.
In our analysis, bearing in mind dominance of the one-photon
emission,
we take into
account
the photon absorption by multiplying the theoretical
electron
struggling function by the averaged one-photon
absorption
factor
$\langle
K_{abs}\rangle\approx1-L/2\lambda_{ph}$,
where $\lambda_{ph}$ is the photon attenuation
length.
This factor decreases our theoretical predictions
by
$\lsim 1-3\%$ for the carbon and aluminum targets used in
\cite{SL2}.
For other targets the effect is even
smaller.

It is appropriate here to comment also about the
status
of Eqs.~(\ref{eq:30}),~(\ref{eq:60})
when
multi-photon
effects become important. It is clear that for $L\gsim
X_{0}$
the formulas of previous section are
inapplicable.
Nonetheless, Eqs.~(\ref{eq:30}),~(\ref{eq:60}) will
give
approximately
right predictions for the intensity of radiation of soft photons
with
$L_{f}^{'}\ll L$ on the targets with $L \ll X_{0}$. In this
case
one can neglect the possibility of radiation of hard
photons,
and
evaluate
the bremsstrahlung rate at $x \ll 1$ ignoring the electron
energy
loss and energy correlations for emitted
photons.
Then, one can easily show that multi-photon contribution
completely
cancels
the effect of the attenuation factor for the one-photon emission,
and
the intensity of bremsstrahlung is given by the formula
obtained
for emission of a single
photon.

\section{Numerical results and comparison with the SLAC data}
We will compare our predictions with the SLAC data
\cite{SL2}
for the
targets
with $L\sim 0.7-6\%\,X_{0}$. We exclude from our analysis the data
for
the gold
target
with $L=0.1\%\,X_{0}$ for which there is a problem with
normalization
of the experimental spectrum
\cite{SL2}.
The measurements of Ref. \cite{SL2} were
performed
with 8 and 25
GeV
electron
beams
for the total radiated energy from 200 keV to 500
MeV.
In the present paper
we
restrict ourselves to the region above 5
MeV.
In this case one can neglect the dielectric
effect
which was not included in our analysis. On the other
hand,
this also
justifies,
as was argued in section 3, the probabilistic treatment of
the
multi-photon effects for the targets with $L\sim
2-6\%\,X_{0}$.
At $E_{e}=8$ GeV this approach can be also used for the
$0.7\%\,X_{0}$
gold target. To illustrate the degree
of
nonlocality in photon emission
we
show in Table 2 the ratio $L/L_{f}^{'}$ at $k=5$ and 100
MeV
for the targets with $L\lsim 3\%\,X_{0}$ used in the SLAC
experiment
\cite{SL2} evaluated using
Eq.~(\ref{eq:234})
(we adopt for the targets
the
notations of
Ref. \cite{SL2}).
Table 2 demonstrates that even for the lower bound of
our
kinematical domain the inequality $L_{f}^{'} \ll L$
is
satisfied for all the targets with $L\gsim 2\%\,X_{0}$,
and
at $E_{e}=8$ GeV also for
the
$0.7\%\,X_{0}$ gold
target.
The only exception is the $0.7\%\,X_{0}$ gold
target
at $E_{e}=25$ GeV. In this case the probabilistic
approach
becomes applicable for $k\gsim 50-100$
MeV.
However, for the
$0.7\%\,X_{0}$
target
the $K$-factor is close to unity. For this reason the
inaccuracy
of the probabilistic approach cannot lead to considerable
errors
in the region $k\lsim 50-100$ MeV, and we will use
the
probabilistic $K$-factor (\ref{eq:310}) in this case as
well.

In Fig.~1 we have plotted the infinite medium
suppression
factor as a function of the photon
momentum
for carbon, aluminum, iron, tungsten and uranium for 8 and 25
GeV
electrons. It is seen that the LPM effect is
considerably
stronger for 25 GeV electrons. The upper bound of the region
of
$k$ where LPM suppression becomes significant for the
results
shown in Fig.~1 agrees with the values of
$k_{LPM}$
given in Table
1.

Our numerical calculations show that at $E_{e}=8$ GeV for all
the
targets used in \cite{SL2} the finite-size effects are
negligible
and the exact suppression factor is close to that for infinite
medium.
At $E_{e}=25$ GeV they become
sizeable
at $k\lsim 10$ MeV for the targets with $L\sim
2-3\%\,X_{0}$.
To illustrate the role of the finite-size effects we
have
plotted in Fig.~2 the results
for
the suppression factor for the $0.7\%\,X_{0}$ gold
and
$2\%\,X_{0}$ lead targets (solid
line).
We also show in this
figure
the results for infinite medium (dashed line) and predictions
of
the frozen-size approximation (dotted line)
(\ref{eq:140}).
One can see from Fig.~2 that at $E_{e}=8$ GeV the
finite-size
effects are negligible. For the lead target at $E_{e}=25$ GeV
at
$k\sim 5$ MeV the edge
effects
increase the radiation rate by $\sim 10\%$
and
become negligible for $k\gsim 15-20$ MeV. In the case of
the
$0.7\%\,X_{0}$ gold target for 25 GeV electrons the exact
suppression
factor differs strongly from that for infinite medium
at
$k\lsim 10-15$ MeV. It is seen that in this region the
frozen-size
approximation works well. The transition between the infinite
medium
and the frozen-size regime occurs at $k\sim 20$
MeV.
In terms of the photon formation
length
(for 25 GeV
electrons
$L_{f}(\mbox{mm})\approx 0.94\cdot (1 \mbox{MeV}/k(\mbox{MeV}))$
in
the region
of
$k$ shown in Fig.~2) the borderline between
the
two regimes is
at
$L_{f}/L\sim 2$ (for the $0.7\%\,X_{0}$ gold target $L=0.023$
mm),
and in terms of the medium-modified formation
length
(\ref{eq:234}) $L_{f}^{'}/L\sim
0.6$.
Thus, Fig.~2b
shows
that the transition between
the
infinite-medium and the frozen-size regime
occurs
in a sufficiently narrow region in the vicinity of $k$
corresponding
to $L_{f}^{'}\sim
L/2$.

To illustrate the role of multi-photon emission
we
show in
Fig.~3
$k$-dependence of the $K$-factor (\ref{eq:310}) for some of
the
targets used in
Ref. \cite{SL2}.
It is seen that multi-photon effects are
important
for the targets with $L\gsim 2\%\,X_{0}$. However, we see that, as
was
above said, they are marginal for the $0.7\%\,X_{0}$
gold
target.
Note that the decrease of the $K$-factor at small $k$ seen
from
Fig.~3 is connected with the $x$-dependence of the
exponential
attenuation factor on the right-hand side of
Eq. (\ref{eq:310}).

In Figs. 4-10 we compare our
predictions
(solid line) with the experimental spectra of
Ref. \cite{SL2}
(the theoretical predictions and experimental
data
presented in the same form as in
Ref. \cite{SL2}).
The theoretical curves have been
obtained
taking into account multi-photon emission and photon
absorption.
To demonstrate the role of the LPM effect better we
also
show the Bethe-Heitler spectrum (dashed line). The theoretical
curves
in Figs.~4-10 were multiplied by the normalization
constants,
$C_{norm}$, which were adjusted to minimize $\chi^{2}$ for
our
predictions.
Their values and the corresponding $\chi^{2}$ per degree of
freedom
are given in Table
3.
For most of the targets the fit quality is quite
good
$\chi^{2}/N\sim 1$ (the value of $\chi^{2}/N$ averaged over all
the
targets is $\sim 1.5$). This says that our predictions
describe
well the shape of the experimental spectra. This
is
also seen directly from
Figs.~4-10.
For
the
$0.7\%\,X_{0}$ gold target at 25 GeV (Fig.~8b) we have also
depicted
the spectrum
obtained
with the infinite medium suppression factor
(dot-dashed
line). It is seen that for this version the curve goes below
the
experimental points
for
$k\lsim 30$
MeV,
while the
curve
obtained including the finite-size effects is close to the
experimental
spectrum in the whole range of
$k$.
It is worth to emphasize that our theoretical predictions do
not
contain
fitting parameters except the normalization
constants.

From Table 3 one can see
that
for most of the targets the agreement of our predictions
with
experimental data in the absolute cross section is
within
$\sim 5\%$. This is close
to
estimate
of the systematic error $\sim 3.5-4.6\%$ given by the authors
of
Ref. \cite{SL2}.
However,
the
disagreement in normalization is rather
big
for
the uranium targets
($C_{norm}\approx
0.86-0.89$ for both 8 and 25 GeV
beams),
and for the $0.7\%\,X_{0}$ gold target for 8 GeV
beam
($C_{norm}\approx
1.17$).
The origin of the above
disagreement
in normalization is not
clear.
Note that the normalization constant
for
the $0.7\%\,X_{0}$ gold target for $E_{e}=25$ GeV obtained
in
the present paper is bigger than that of
our
previous analysis \cite{LPM2} by $\sim
13\%$.
This discrepancy is connected
with
neglecting
the Coulomb correction to the dipole cross section and
multi-photon
emission in Ref. \cite{LPM2} which, however, practically do not
affect
the shape of the the
spectrum.

\section{Conclusion}
In the present paper we have carried
out
a detailed
theoretical
analysis
of
the recent SLAC data \cite{SL2} on the LPM
effect
for bremsstrahlung from 8 and 25 GeV electrons in a variety
of
materials. The calculations have been performed within a
rigorous
light-cone path
integral
approach to the LPM effect previously developed in
Refs.
\cite{LPM1,LPM2}, which reduces evaluation of the radiation rate
to
solution of a
two-dimensional
Schr\"odinger
equation with an imaginary potential. This potential is
proportional
to the dipole cross section for scattering of $e^{+}e^{-}$
pair
off an
atom.
In our calculations we treat rigorously the
Coulomb
effects and include the inelastic
processes.
We have compared the theoretical prediction with the SLAC data
taking
into account multi-photon emission and photon
absorption.
For most of the targets our predictions are in very good
agreement
with the experimental data. In particular, we describe
well
the spectrum for the $0.7\%\,X_{0}$ gold target at $E_{e}=25$
GeV
for which the finite-size effects play an important
role.\\
\bigskip
\phantom\\
{\large \bf
Acknowledgements}

I would like to thank S.R. Klein for sending the files of
the
experimental data taken by the E-146 collaboration and
helpful
comments about some aspects of the SLAC
experiment.
This work was partially supported by the
INTAS
grants 93-239ext and
96-0597.

\newpage

\begin{table}[h]
\caption[.]{ The values of $k_{LPM}$  obtained
using Eq.~(\ref{eq:233p}) for the target materials used in the
SLAC
experiment \cite{SL2}. In the third column we give
the
radiation lengths used in Ref. \cite{SL2} and in the present
paper.}  
\begin{center}
\begin{tabular}{||c||c|c|c|c|} \hline\hline
 Material & $Z$ & $X_{0}$ & $k_{LPM}$(MeV) & $k_{LPM}$(MeV)
           \\
           & &(mm) & ($E_{e}=8$ GeV) & ($E_{e}=25$ GeV) \\
           \hline
           C & 6 & 196 & 0.42 & 4.1
           \\
           Al & 13 & 89 & 0.93 & 9.1
           \\
           Fe & 26 & 17.6 & 4.7 & 46.1
           \\
           W & 74 & 3.5 & 23.8& 230
           \\
           Au & 79 & 3.3 & 25.1 & 244
           \\
           Pb & 82 & 5.6 & 14.8 & 144
           \\
           U & 92 & 3.5 & 23.7 & 230
           \\
\hline\hline
\end{tabular}
\end{center}
\end{table}

\newpage

\begin{table}[h]
\caption[.]{ The ratio $L/L_{f}^{'}$ for the
targets with $L\lsim 3\%\,X_{0}$ used in
Ref. \cite{SL2}
estimated with the help of
Eq.~(\ref{eq:234}).}  
\begin{center}
\begin{tabular}{||c||c|c|c|c|} \hline\hline
 Target & \multicolumn{2}{c|}{$E_{e}=8$ GeV}
  &
  \multicolumn{2}{c|}{$E_{e}=25$ GeV}
  \\
  \cline{2-5}
  & $k=5$ MeV
  &
  $k=100$ MeV & $k=5$ MeV
  &
  $k=100$ MeV \\
  \hline
  2\%C & 213 & 4270 & 22 & 437
  \\
  3\%Al & 162 & 3250 & 22 & 333
  \\
  3\%Fe & 25.5 & 510 & 8 & 52
  \\
  2\%W & 10 & 92 & 3.3 & 15
  \\
  0.7\%Au & 2.7 & 24 & 0.86 & 3.85
  \\
  2\%Pb & 13.5 & 156 & 4.3 & 19.3
  \\
  3\%U & 9 & 82 & 2.8 & 12.6
  \\
\hline\hline
\end{tabular}
\end{center}
\end{table}

\newpage

\begin{table}[h]
\caption[.]{ List of the normalization constants adjusted to
match our predictions (solid line in
Figs.~4-10)
with the SLAC data
\cite{SL2}.
The
corresponding
$\chi^{2}$ per degree of freedom are also
given.
The second column gives the target thicknesses in
mm.}  
\begin{center}
\begin{tabular}{||c||c|c|c|c|c|} \hline\hline
 Target & $L$ & \multicolumn{2}{c|}{$E_{e}=8$ GeV}
  &
  \multicolumn{2}{c|}{$E_{e}=25$ GeV}
  \\
  \cline{3-6}
  & (mm) & $C_{norm}$ & $\chi^{2}/N$ & $C_{norm}$ & $\chi^{2}/N$
  \\
  \hline
  2\%C & 4.1 &$0.943\pm 0.004$ & 0.98 &$0.957\pm 0.003$ & 4.2
  \\
  6\%C & 11.7 &$0.964\pm 0.004$ & 1.14 &$0.964\pm 0.002$ & 3.46
  \\
  3\%Al & 3.12 &$0.985\pm 0.003$ & 1.02 &$0.981\pm 0.003$ & 1.41
  \\
  6\%Al & 5.3 & & &$0.982\pm 0.003$ & 1.6
  \\
  3\%Fe & 0.49 &$1.00\pm 0.005$ & 0.79 &$0.972\pm 0.002$ & 1.85
  \\
  6\%Fe & 1.08 & & &$0.96\pm 0.002$ & 1.55
  \\
  2\%W & 0.088&$0.942\pm 0.003$ & 1.14 &$0.953\pm 0.003$ & 2.8
  \\
  6\%W & 1.08 & & &$1.007\pm 0.003$ & 1.45
  \\
  0.7\%Au & 0.023 &$1.174\pm 0.007$ & 1.44 &$1.056\pm 0.004$ & 0.8
  \\
  6\%Au & 0.2 & $1.014\pm 0.003$& 1.15 &$1.031\pm 0.002$ & 0.89
  \\
  2\%Pb & 0.15 & $1.032\pm 0.004$& 1.01 &$1.009\pm 0.002$ & 0.94
  \\
  3\%U & 0.079 & $0.875\pm 0.003$& 1.0 &$0.886\pm 0.002$ & 2.63
  \\
  5\%U & 0.147 & $0.865\pm 0.004$& 1.04 &$0.877\pm 0.003$ & 1.63
  \\
\hline\hline
\end{tabular}
\end{center}
\end{table}

\newpage

\begin{center}
{\Large \bf Figures}
\end{center}

\begin{figure}[h]
\begin{center}
\epsfig{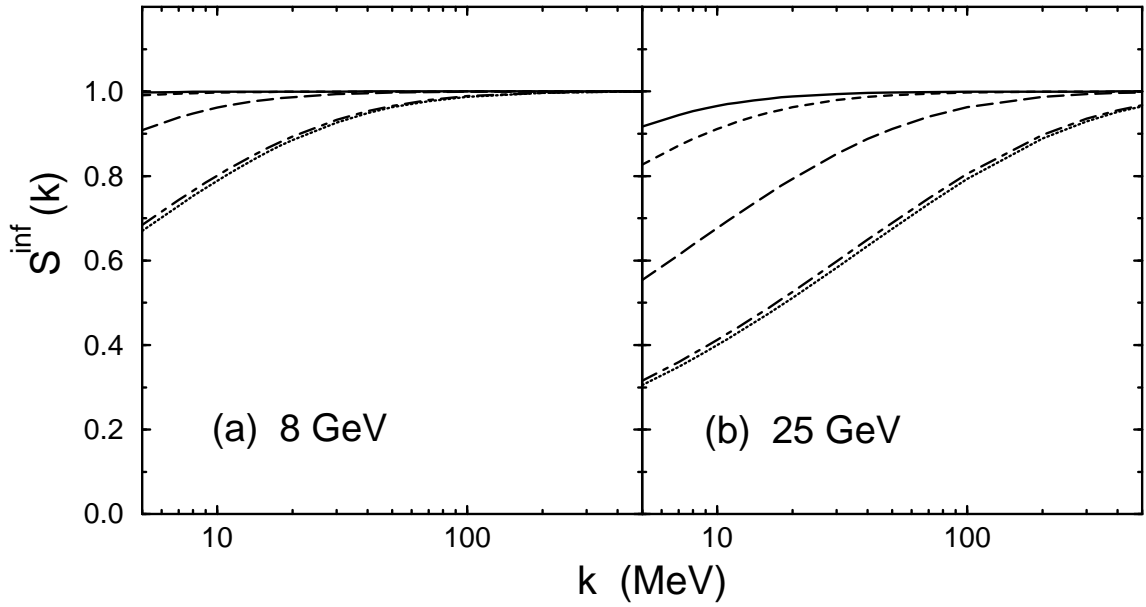}
\end{center}
\caption[.]{
The infinite medium suppression factor
for
bremsstrahlung from 8 (a) and 25 (b) GeV
electrons
for carbon (solid line), aluminum (dashed line), iron
(long-dashed
line), tungsten (dot-dashed line) and uranium (dotted
line).
}
\end{figure}

\pagebreak

\begin{figure}[t]
\begin{center}
\epsfig{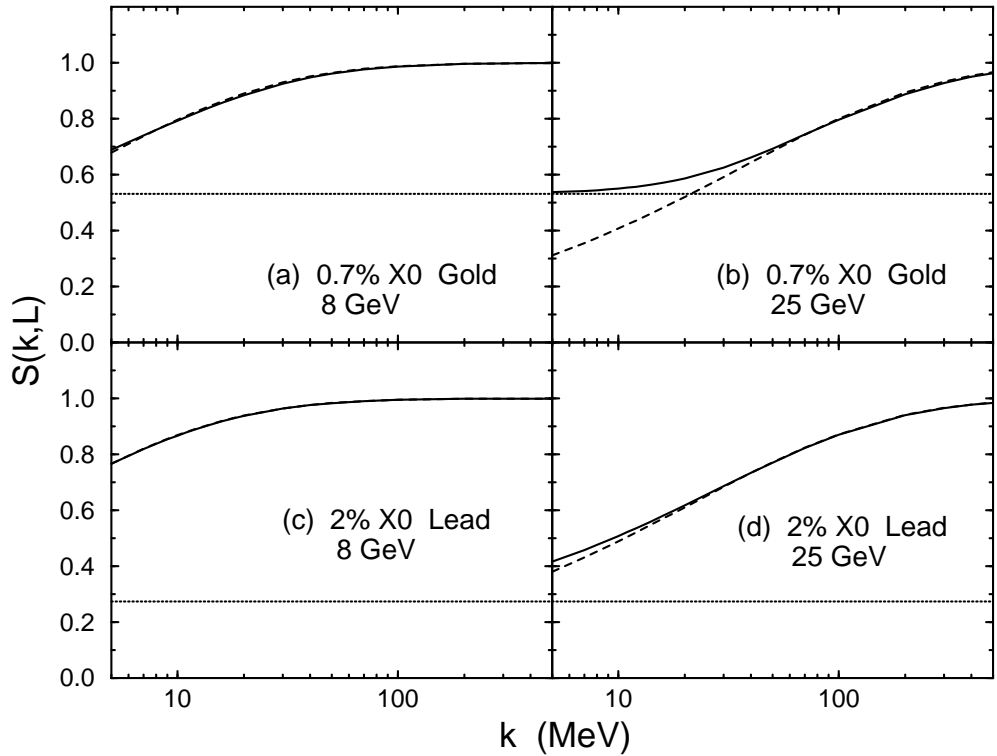}
\end{center}
\caption[.]{
The LPM suppression factor for the $0.7\%\,X_{0}$ gold
and
the $2\%\,X_{0}$ lead targets (solid line). The dashed line
shows
the results for infinite medium, and the dotted line
corresponds
to the frozen-size approximation 
(\ref{eq:140}).}
\end{figure}


\begin{figure}[h]
\begin{center}
\epsfig{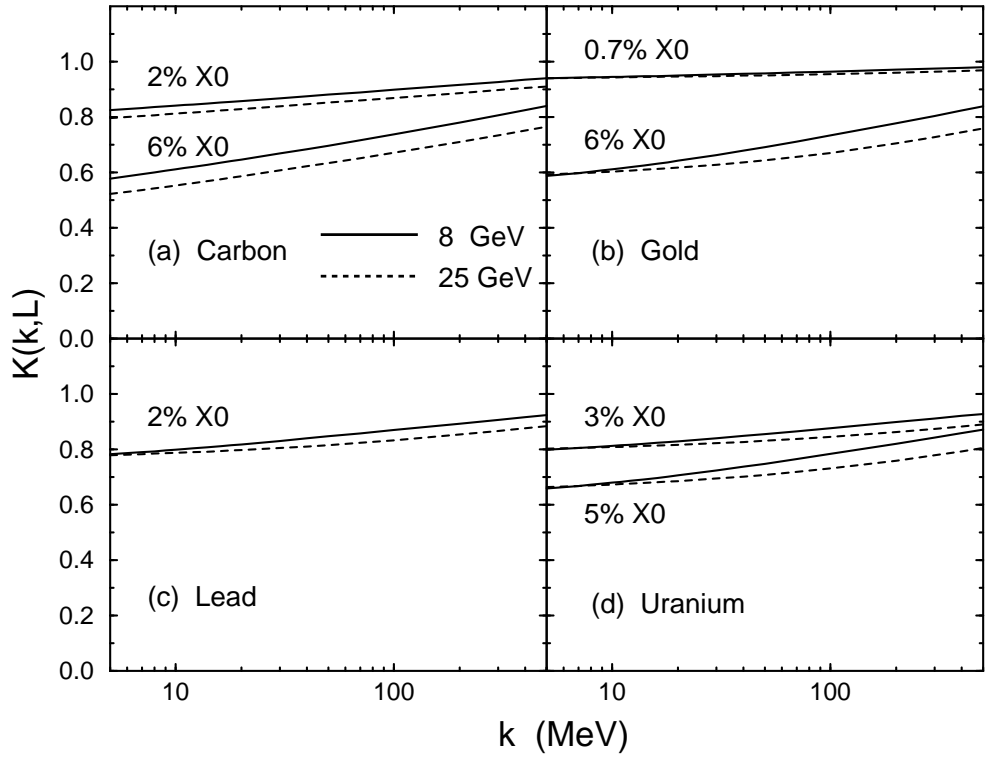}
\end{center}
\caption[.]{
The multi-photon $K$-factor
(\ref{eq:310})
for the carbon (a), gold (b), lead (c) and uranium (d)
targets.
The solid and dashed line correspond to 8 and 25 beams,
respectively.}
\end{figure}

\pagebreak

\begin{figure}[t]
\begin{center}
\epsfig{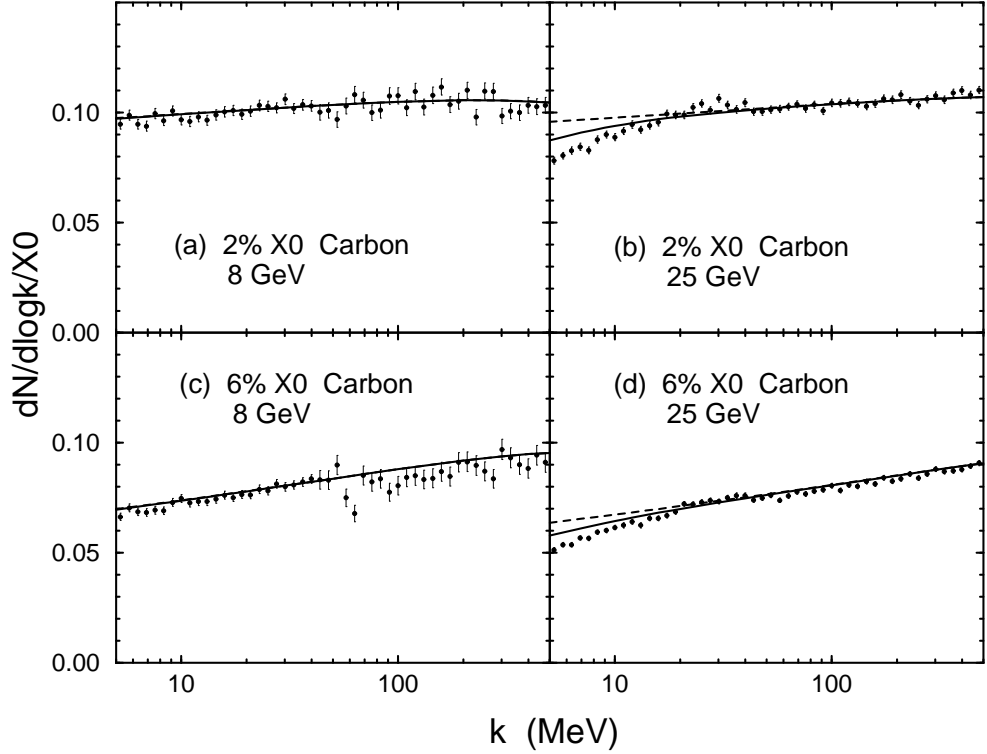}
\end{center}
\caption[.]{
The spectrum in the radiated energy for the $2\%\,X_{0}$
(a,~b)
and $6\%\,X_{0}$ (c,~d) carbon targets. The experimental
data
from Ref. \cite{SL2}. The solid line shows our results
obtained
using Eq.~(\ref{eq:60}). The dashed line shows
the
Bethe-Heitler spectrum. In both these cases the
multi-photon
emission and photon absorption are taken into
account.
}
\end{figure}

\pagebreak

\begin{figure}[t]
\begin{center}
\epsfig{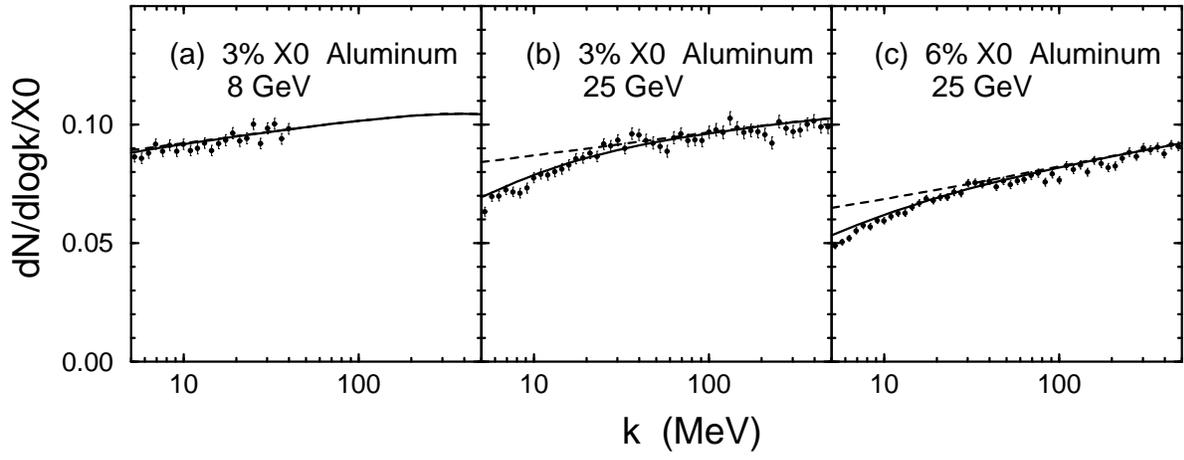}
\end{center}
\caption[.]{
The same as in Fig.~4 but
for
the $3\%\,X_{0}$ (a,~b) and $6\%\,X_{0}$ (c) aluminum
targets.
}
\end{figure}


\begin{figure}[h]
\begin{center}
\epsfig{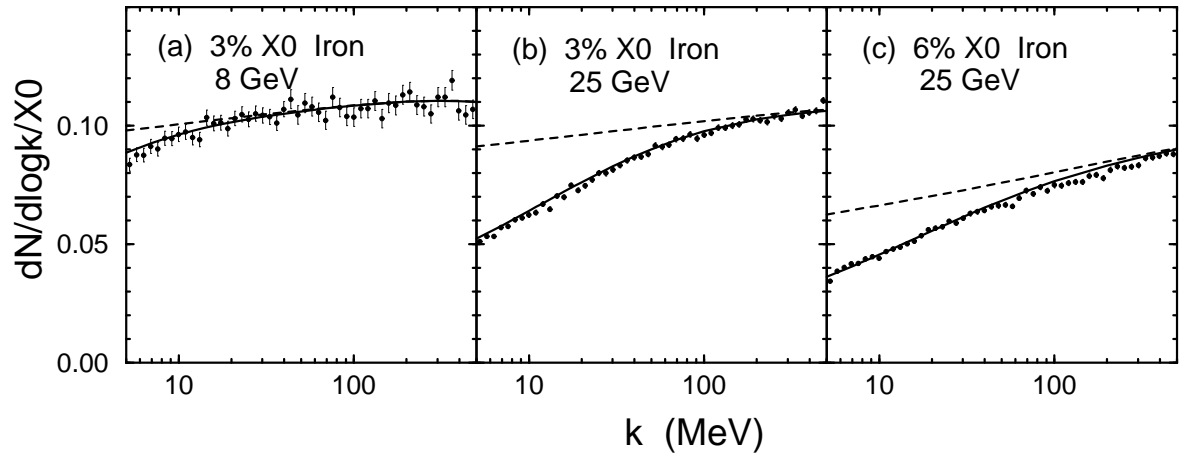}
\end{center}
\caption{
The same as in Fig.~4 but
for
the $3\%\,X_{0}$ (a,~b) and $6\%\,X_{0}$ (c) iron
targets.
}
\end{figure}

\pagebreak

\begin{figure}[t]
\begin{center}
\epsfig{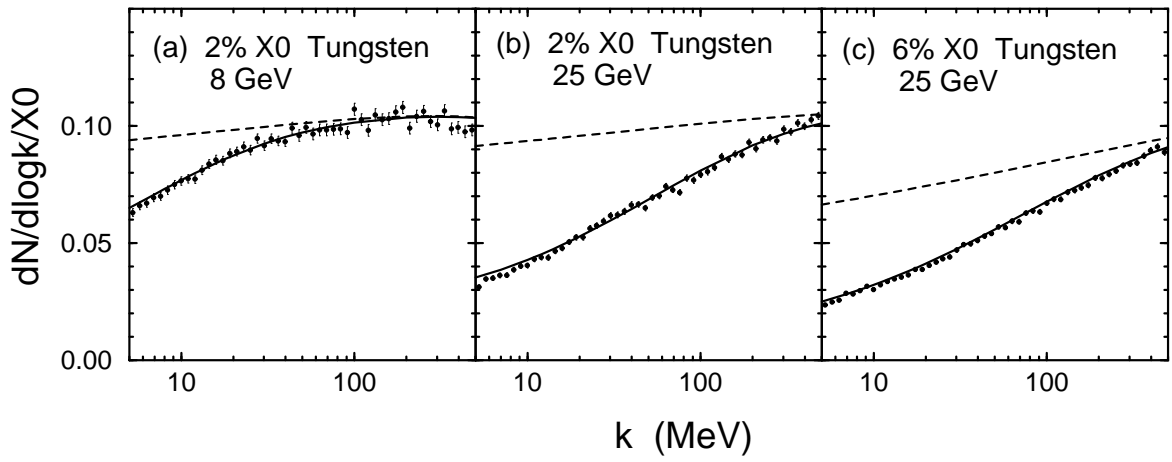}
\end{center}
\caption{
The same as in Fig.~4 but
for
the $2\%\,X_{0}$ (a,~b) and $6\%\,X_{0}$ (c) tungsten
targets.
}
\end{figure}


\begin{figure}[b]
\begin{center}
\epsfig{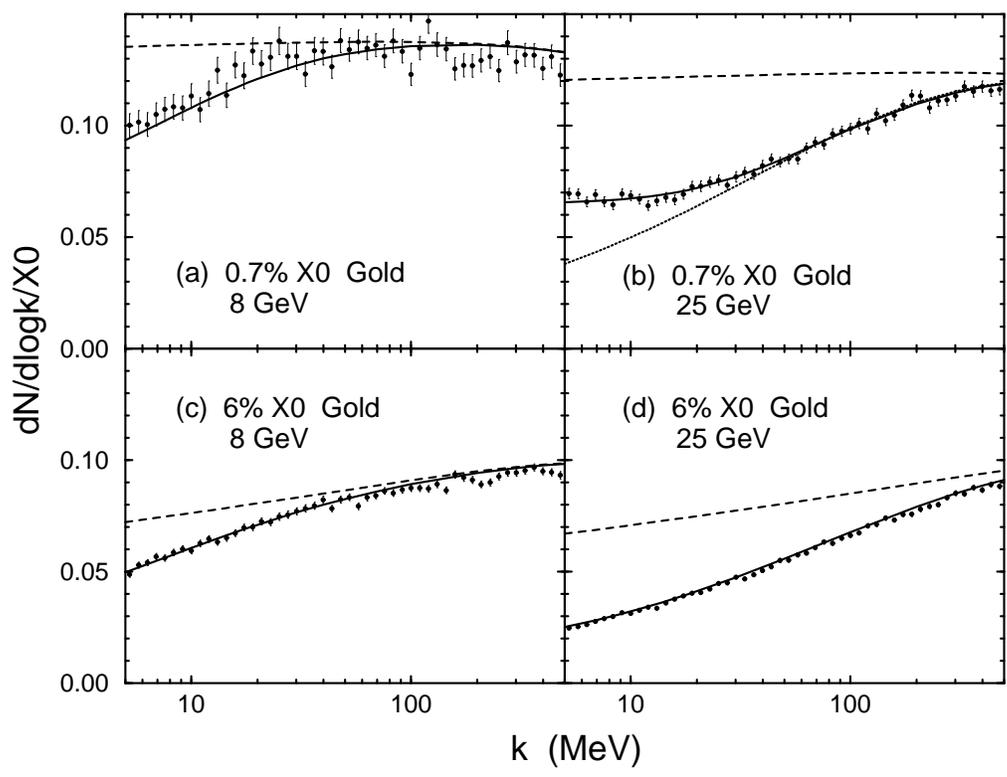}
\end{center}
\caption{
The same as in Fig.~4 but
for
the $0.7\%\,X_{0}$ (a,~b) and $6\%\,X_{0}$ (c,~d) gold
targets.
}
\end{figure}

\pagebreak

\begin{figure}[t]
\begin{center}
\epsfig{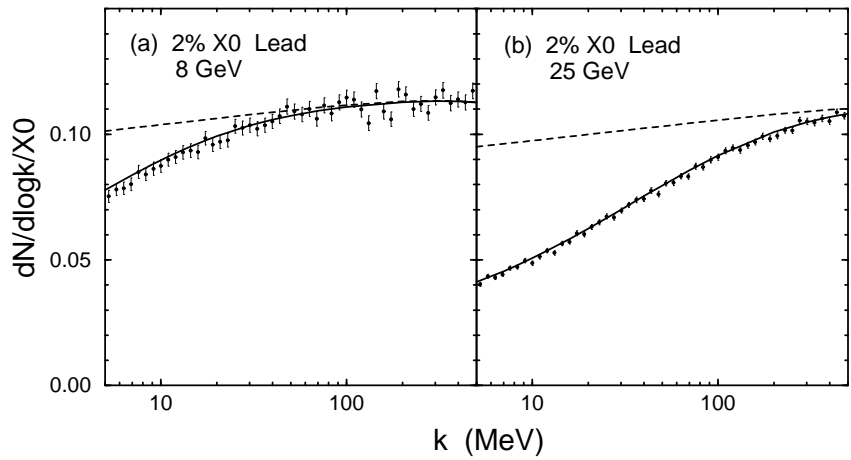}
\end{center}
\caption{
The same as in Fig.~4 but
for
the $2\%\,X_{0}$ lead
target.
}
\end{figure}


\begin{figure}[b]
\begin{center}
\epsfig{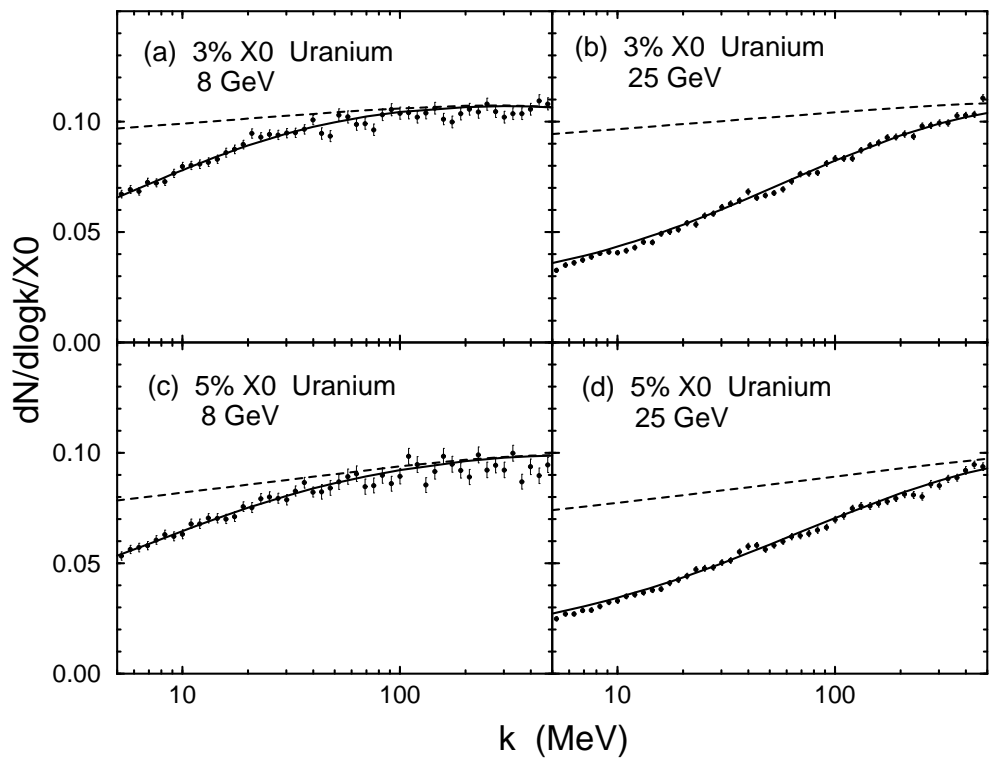}
\end{center}
\caption{
The same as in Fig.~4 but
for
the $3\%\,X_{0}$ (a,~b) and $5\%\,X_{0}$ (c,~d) uranium
targets.
}
\end{figure}


\newpage
\begin{thebibliography}{99}

\bibitem{LP}
L.D. Landau and
I.Ya. Pomeranchuk,
{\sl Dokl. Akad. Nauk SSSR} {\bf 92} (1953) 535,
735.

\bibitem{Migdal}
A.B. Migdal, {\sl Phys. Rev.} {\bf 103} (1956)
1811.

\bibitem{TM}
M.L. Ter-Mikaelian, {\sl High Energy Electromagnetic Processes
in
Condensed Media} (Wiley, NY,
1972).

\bibitem{Ryazanov}
M.I. Ryazanov, {\sl Sov. Phys. Usp.} {\bf
17}
(1975)
815.

\bibitem{Akh}
A.I. Akhiezer and N.F. Shul'ga, {\sl Sov. Phys. Usp.} {\bf
30}
(1987)
197.

\bibitem{Klein}
S.R. Klein, {\sl Preprint} {\bf LBNL-41350}, Berkeley
(1998);
to appear in
{\sl Rev. Mod. Phys..}

\bibitem{Serp}
A.A.Varfolomeev, V.I.Glebov, E.I.Denisov {\sl et
al}.,
{\sl Sov. JETP} {\bf 42} (1975)
218.

\bibitem{C1}
T.Stanev, Ch.Vankov, R.E.Streitmatter {\sl et
al}.,
{\sl Phys. Rev.} {\bf D25} (1982)
1291.

\bibitem{C2}
K.Kasahara,
{\sl Phys. Rev.} {\bf D31} (1985)
2737.

\bibitem{SL1}
P.L. Anthony, R. Becker-Szendy, P.E. Bosted {\sl et
al}.,
{\sl Phys. Rev. Lett.} {\bf 75} (1995)
1949.

\bibitem{SL2}
P.L. Anthony, R. Becker-Szendy, P.E. Bosted {\sl et
al}.,
{\sl Phys. Rev.} {\bf D56} (1997)
1373.

\bibitem{ShF}
N.F.Shul'ga and S.P. Fomin, {\sl JETP Lett.} {\bf 63} (1996)
873.

\bibitem{BlanD}
R. Blankenbecler and S.D. Drell, {\sl Phys. Rev.} {\bf D53} (1996)
6265.

\bibitem{LPM1}
B.G. Zakharov, {\sl JETP Lett.} {\bf 63} (1996)
952.

\bibitem{B2}
R. Baier, Yu.L. Dokshitzer, A.H. Mueller, S. Peigne and
D. Schiff,
{\sl Nucl. Phys.}
{\bf B478} (1996)
577.

\bibitem{LPM2}
B.G. Zakharov, {\sl JETP Lett.} {\bf 64} (1996)
781.

\bibitem{Blan}
R. Blankenbecler, {\sl Phys. Rev.} {\bf D55} (1997)
190.

\bibitem{A2eBGZ}
B.G. Zakharov, {\sl Sov. J. Nucl. Phys.} {\bf 46} (1987)
92.

\bibitem{NPZ}
N.N. Nikolaev, G.Piller and
B.G. Zakharov,
{\sl JETP} {\bf 81} (1995)
851.

\bibitem{Coulomb}
H.Davies, H.A.Bethe and L.C.Maximon, {\sl Phys. Rev.} {\bf 93}
(1954)
788.

\bibitem{Tsai}
Y.-S. Tsai, {\sl  Rev. Mod. Phys.} {\bf 46} (1974)
815.

\bibitem{Rossi}
B.Rossi and K.Greisen, {\sl Rev. Mod. Phys.} {\bf 13} (1941)
240.

\bibitem{Eyges}
L.Eyges, {\sl Phys. Rev.} {\bf 70} (1949)
264.

\bibitem{BHstrug}
H.A.Bethe and W.Heitler, {\sl Proc. Roy. Soc.} {\bf A146} (1934)
83.

\end{thebibliography}
\end{document}